\documentclass[twocolumn,10pt]{article}
\usepackage[margin=0.8in]{geometry}
\usepackage{amsmath,amssymb,graphicx,bm}
\usepackage{braket}
\usepackage[colorlinks=true,linkcolor=blue,citecolor=blue,urlcolor=blue]{hyperref}
\usepackage{authblk}
\usepackage{xcolor}
\usepackage{tikz}
\usetikzlibrary{arrows.meta,decorations.pathmorphing,decorations.markings,positioning,calc,fit,backgrounds,shapes.geometric,patterns,shapes.misc}

\definecolor{victim}{HTML}{1F4E79}
\definecolor{advers}{HTML}{B03A2E}
\definecolor{buffer}{HTML}{909497}
\definecolor{accent}{HTML}{B7950B}
\definecolor{faint}{HTML}{D5DBDB}
\definecolor{coh}{HTML}{B03A2E}
\definecolor{sto}{HTML}{1F4E79}

\newcommand{\Tr}{\operatorname{Tr}}
\newcommand{\dnorm}[1]{\left\lVert #1 \right\rVert_{\diamond}}
\newcommand{\tnorm}[1]{\left\lVert #1 \right\rVert_{1}}
\newcommand{\Leak}{\mathcal{L}}
\newcommand{\Chan}{\mathcal{E}}
\newcommand{\Xset}{\mathcal{X}}
\newcommand{\defeq}{\triangleq}

\title{Coherence Rather Than Error Rate Governs Privacy in Multi-Tenant Quantum Computing}

\author[1]{Farhad~Farokhi\thanks{Corresponding author: farhad.farokhi@unimelb.edu.au}}
\affil[1]{\small Department of Electrical and Electronic Engineering, The University of Melbourne, Australia}
\date{\small \today}

\begin{document}
\twocolumn[
\begin{@twocolumnfalse}
\maketitle

\begin{abstract}
Multi-tenant computing enables providers of commercial cloud quantum processors to rent disjoint sectors of a device to independent users. Average gate error, which cloud quantum computing providers report, does not determine how much one tenant learns about another. We propose an information-theoretic notion of information leakage across co-tenancy boundaries stemming from quantum state distinguishability. We measure this leakage on commercially-available $156$-qubit (IBM Kingston) and $20$-qubit (IQM Garnet) devices. 
Boundaries with identical benchmarked error can offer significantly different amount of information leakage because standard reported measures of error are blind to coherent-versus-stochastic nature of  the error while the proposed notion of information leakage is not. A uniform Pauli randomisation implemented over the victim's whole register is used as a defence mechanism to reduce the information leakage to zero. The defence theoretically does not incur a fidelity cost, but the experiments show a non-trivial degradation caused by accumulation of errors. We provide a specific call-for-action to the providers of quantum cloud computing to report information leakage in addition to standard error rates in their device datasheet to enable users to compute privacy and security risks prior to engagement with the device.
\end{abstract}
\vspace{1em}
\end{@twocolumnfalse}
]

\section*{Introduction}
Quantum processors are expensive. Cloud interfaces, where one can rent portions of quantum processors, are how most people use them. Useful circuits are small compared to the devices that run them. This motivates the providers of cloud quantum computing services to put more than one user on the same device at once~\cite{ref:multiprog}, albeit each in a separate sector.

Sharing a quantum computer, however, has a cost. Adjacent sectors are never perfectly isolated. This residual coupling, called crosstalk, is well characterised and providers already work to reduce it~\cite{ref:sarovar,ref:zhou}. However, crosstalk is an information hazard as well as an error term. The same mechanism that degrades a tenant's gate fidelity, in one direction, carries information about that tenant, in the other direction, by unintentionally entangling qubits across the boundary.

The information leakage has been documented. Crosstalk signatures reveal a neighbour's gate structure and can classify their algorithm \cite{ref:choudhury}. When using SWAP operations to implement gates between logically distant qubits, the whole chain of operations leaks information~\cite{ref:swapattack, 1011453708821}. Shared readout hardware also leaks information~\cite{ref:readout}. Defences based on dynamical decoupling have been proposed~\cite{ref:mehra,ref:harper}. One recent study models this leakage with a learned decoder and experimentally observes the relationship between degrading the neighbour's fidelity and information leakage, but it does not quantitatively connect the two effects~\cite{ref:bell}. 

We ask a question that is relevant to providers of cloud quantum processors. Can a tenant compute privacy guarantees from the statistics that the provider already publishes, such as the device datasheet? The answer is no. The average gate infidelity, a statistic the providers report, does not distinguish a coherent error from a stochastic one. We show information leakage depends almost entirely on this characteristic. A low reported error rate can coincide with a large privacy leakage. Information leakage can be eliminated by the existing randomised circuit compiling techniques. The results of this paper provide a specific call-for-action: the providers of cloud quantum computing should report information leakage in addition to standard error rates in their device datasheet to enable users to compute risks of information leakage prior to engagement with the device. 

\section*{Results}
\subsection*{Setting.}

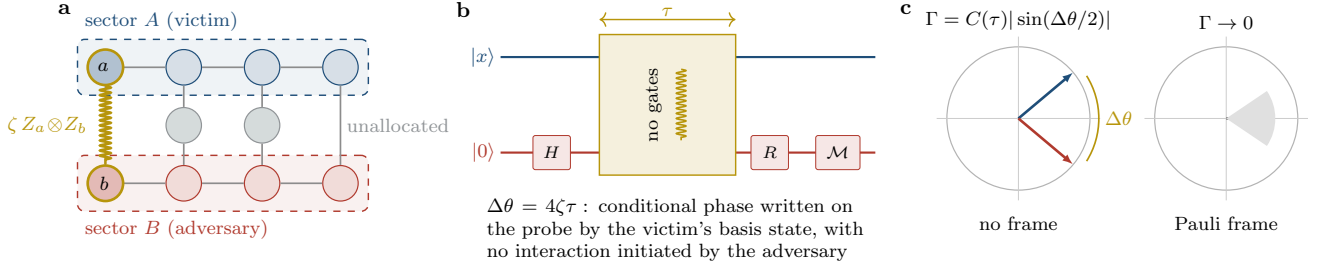
\begin{figure*}[t]
\resizebox{\textwidth}{!}{
\begin{tikzpicture}[
  font=\small,
  qubit/.style={circle,draw=black!70,line width=0.5pt,minimum size=5.2mm,
                inner sep=0pt,fill=white},
  vq/.style={qubit,fill=victim!18,draw=victim},
  aq/.style={qubit,fill=advers!18,draw=advers},
  bq/.style={qubit,fill=faint,draw=buffer},
  cpl/.style={draw=black!45,line width=0.7pt},
  lbl/.style={font=\footnotesize},
  pan/.style={font=\bfseries\small}
]

\begin{scope}[shift={(0,0)}]
\node[pan] at (-0.6,2.55) {a};
\foreach \i in {0,...,3}{
  \node[vq] (v\i) at (\i*1.15,1.7) {};
  \node[aq] (a\i) at (\i*1.15,0.0) {};
}
\foreach \i in {0,...,2}{
  \draw[cpl] (v\i) -- (v\the\numexpr\i+1\relax);
  \draw[cpl] (a\i) -- (a\the\numexpr\i+1\relax);
}
\foreach \i in {0,3}{ \draw[cpl] (v\i) -- (a\i); }
\node[bq] (b1) at (1.15,0.85) {};
\node[bq] (b2) at (2.30,0.85) {};
\draw[cpl] (v1) -- (b1) -- (a1);
\draw[cpl] (v2) -- (b2) -- (a2);
\begin{scope}[on background layer]
  \node[fit=(v0)(v3),draw=victim,dashed,line width=0.6pt,rounded corners=3pt,
        inner sep=4pt,fill=victim!7] (boxV) {};
  \node[fit=(a0)(a3),draw=advers,dashed,line width=0.6pt,rounded corners=3pt,
        inner sep=4pt,fill=advers!7] (boxA) {};
\end{scope}
\node[lbl,victim,anchor=west] at ($(boxV.north west)+(0,0.26)$) {sector $A$ (victim)};
\node[lbl,advers,anchor=west] at ($(boxA.south west)+(0,-0.26)$) {sector $B$ (adversary)};
\node[lbl,buffer,anchor=west] at ($(b2.east)+(0.85,0)$) {unallocated};
\node[vq,draw=accent,line width=1.1pt,fill=victim!35] (va) at (0,1.7) {};
\node[aq,draw=accent,line width=1.1pt,fill=advers!35] (pb) at (0,0.0) {};
\node[lbl,anchor=east] at ($(va.west)+(+0.45,0)$) {$a$};
\node[lbl,anchor=east] at ($(pb.west)+(+0.45,0)$) {$b$};
\draw[accent,line width=1.0pt,decorate,
      decoration={coil,aspect=0,segment length=2.1pt,amplitude=2.3pt}]
      (va) -- (pb);
\node[lbl,accent,anchor=west] at (-1.56,0.85) {$\zeta\,Z_a\!\otimes\!Z_b$};
\end{scope}

\begin{scope}[shift={(5.8,0)}]
\node[pan] at (-0.55,2.55) {b};
\def\ytop{1.85}\def\ybot{0.45}
\draw[victim,line width=0.9pt] (0,\ytop) -- (5.5,\ytop);
\draw[advers,line width=0.9pt] (0,\ybot) -- (5.5,\ybot);
\node[lbl,victim,anchor=east] at (0.05,\ytop) {$\ket{x}$};
\node[lbl,advers,anchor=east] at (0.05,\ybot) {$\ket{0}$};
\node[draw=advers,fill=advers!12,rounded corners=1pt,minimum width=5.5mm,minimum height=5mm,font=\scriptsize] (H) at (0.75,\ybot) {$H$};
\node[draw=advers,fill=advers!12,rounded corners=1pt,minimum width=5.5mm,minimum height=5mm,font=\scriptsize] (S) at (3.95,\ybot) {$R$};
\node[draw=advers,fill=advers!12,rounded corners=1pt,minimum width=6.5mm,minimum height=5mm,font=\scriptsize] (M) at (4.95,\ybot) {$\mathcal{M}$};
\fill[accent!14] (1.45,\ybot-0.33) rectangle (3.45,\ytop+0.33);
\draw[accent,line width=0.6pt] (1.45,\ybot-0.33) rectangle (3.45,\ytop+0.33);
\draw[<->,accent,line width=0.6pt] (1.45,\ytop+0.55) -- (3.45,\ytop+0.55)
      node[midway,above,lbl,accent,inner sep=1pt] {$\tau$};
\node[lbl,anchor=center,rotate=90] at (2.25,1.15) {no gates};
\draw[accent,line width=0.7pt,decorate,
      decoration={coil,aspect=0,segment length=2.1pt,amplitude=2.0pt}]
      (2.65,\ytop-0.18) -- (2.65,\ybot+0.18);
\node[lbl,anchor=north,text width=5.6cm,align=left] at (2.6,\ybot-0.55) {$\Delta\theta=4\zeta\tau$\;: conditional phase written on the probe by the victim's basis state, with no interaction initiated by the adversary};
\end{scope}

\begin{scope}[shift={(13.4,0.95)}]
\node[pan] at (-1.65,1.6) {c};
\foreach \dx/\ttl/\open in {0/{no frame}/1, 3.05/{Pauli frame}/0}{
  \begin{scope}[shift={(\dx,0)}]
    \draw[black!30,line width=0.5pt] (0,0) circle (1.05);
    \draw[black!20,line width=0.4pt] (-1.2,0) -- (1.2,0);
    \draw[black!20,line width=0.4pt] (0,-1.2) -- (0,1.2);
    \node[lbl,anchor=north] at (0,-1.32) {\ttl};
    \ifnum\open=1
      \draw[victim,line width=1.1pt,-{Latex[length=1.6mm]}] (0,0)
            -- (40:1.05);
      \draw[advers,line width=1.1pt,-{Latex[length=1.6mm]}] (0,0)
            -- (-40:1.05);
      \draw[accent,line width=0.7pt] (32:1.18) arc (32:-32:1.18);
      \node[lbl,accent,anchor=west] at (0:1.15) {$\Delta\theta$};
      \node[lbl,anchor=south,align=center] at (0,1.18)  {$\Gamma=C(\tau)|\sin(\Delta\theta/2)|$};
    \else
      \draw[black!55,line width=1.1pt,-{Latex[length=1.6mm]}] (0,0) -- (0:0.62);
      \fill[black!12] (0,0) -- (34:0.72) arc (34:-34:0.72) -- cycle;
      \node[lbl,anchor=south,align=center] at (0,1.18) {$\Gamma\to0$};
    \fi
  \end{scope}
}
\end{scope}

\end{tikzpicture}
}
\caption{Cross-tenant information leakage on a shared quantum processor. (a) Two tenants occupy disjoint sectors. Coupled qubits $a$ and $b$ in the two disjoint sectors meet across the {tenant boundary}, where a residual $ZZ$ interaction of rate $\zeta$ entangles them unintentionally.  (b) The victim holds a computational-basis state on $a$. The adversary, on $b$, prepares $\ket{+}$, idles for a window $\tau$, then measures in $X$ and $Y$ basis. The adversary starts no interaction and needs no special privilege beyond co-tenancy. (c) Without a defence the two conditional probe states, i.e., the states of the adversary conditioned on the content of victim's qubit, separate by an angle $\Delta\theta = 4\zeta\tau$, giving a distinguishability $\Gamma = C(\tau)|\sin(\Delta\theta/2)|$ with $C(\tau)$ capturing the effect of decoherence on the adversary's qubit. A randomising defence on the victim, called Pauli frame and defined formally in Result~2, collapses the distinguishability while leaving the average gate infidelity theoretically unchanged.}
\label{fig:threat}
\end{figure*}

\emph{Threat model.} The adversary is an ordinary co-tenant, nothing more, i.e., it does not need privileged access. It occupies a sector (a group of qubits) disjoint from the victim's sector. The adversary  submits its own circuits through the same cloud interface the victim uses. It can only read out its own qubits. It cannot touch, measure, or delay the victim's qubits. It has no pulse-level or calibration-level access. It does not learn how the victim randomises its own circuits. It knows the device layout and the published calibration numbers, both public. It may hold extra qubits as ancilla states and entangle its probe state with them before the boundary interaction. 

We split the full register into a victim sector $A$, an adversary sector $B$, and an optional unused, unallocated buffer. Figure~\ref{fig:threat}(a) depicts this in the case of our experiments. The victim holds a classical (random variable) secret $x$, drawn randomly from a finite set $\Xset$ of size $M$. This secret changes what the victim does, i.e., it changes the content of its qubits and/or the implemented circuit. The whole device evolves under a potentially secret-dependent Hamiltonian given by
\begin{equation}
  H^{(x)}(t) = H_A^{(x)}(t)\otimes I + I \otimes H_B(t) + H_{\times}^{(x)}(t).
  \label{eq:ham}
\end{equation}
Here, $H_A^{(x)}(t)$ is the victim's own Hamiltonian acting only on $A$. It may depend on the secret $x$ because the victim's circuit can depend on the secret. The adversary's own Hamiltonian, which only acts on $B$, is denoted by $H_B(t)$. It does not depend on $x$ because the adversary's circuit cannot depend on a secret it does not know. The boundary or cross Hamiltonian term, the only piece that connects the two sectors, is 
\begin{equation}
  H_{\times}^{(x)}(t) = \sum_k \zeta_k^{(x)}(t)\, O_A^{(k)} \otimes O_B^{(k)},\label{eqn:cross}
\end{equation}
where $O_A^{(k)}$ and $O_B^{(k)}$ are local operators, such as Pauli operators, on the boundary qubits of $A$ and $B$, respectively, and $\zeta_k^{(x)}$ is the coupling strength, taking the form of an angular frequency. We quote $\zeta/2\pi$ in Hertz throughout. For experiments in this paper, which investigate the impact of $ZZ$ crosstalk between the qubits, the cross Hamiltonian reduces to one term with $O_A^{(k)} = Z_a$ and $O_B^{(k)} = Z_b$, where $Z_a,Z_b$ denote local Pauli operators. This is depicted in Figure~\ref{fig:threat}(a). 

Tracing the victim out of the picture gives the adversary's conditional channel, $\Chan_B^x$, namely $B$'s evolution for one fixed value of the secret $x$ over time window of $[t,t+\tau]$. The conditional channel satisfies
\begin{align}
    \Chan_B^x(\rho_B^{\mathrm{in}})=
    \Tr_A\left[U(t,\tau)(\rho_A^{\mathrm{in}(x)}\otimes\rho_B^{\mathrm{in}})U(t,\tau)^\dag\right],
\end{align}
where $\rho_B^{\mathrm{in}}$ is any input state the adversary chooses to prepare on its own qubits at time $t$, $\rho_A^{\mathrm{in}(x)}$ is the victim's initial state which may depend on secret $x$ at time $t$, and $U(t,\tau)$ denotes the evolution of the joint system under the Hamiltonian in Eq.~\eqref{eq:ham} over the interval $[t,t+\tau]$. The conditional channel captures whatever happens to the adversary's qubits as a function of the victim's secret $x$ due to the cross Hamiltonian. The state that the adversary can measure is $\rho_B^x = \Chan_B^x(\rho_B^{\mathrm{in}})$. In the experimental hardware measurement described later, the adversary's chosen input is $\rho_B^{\mathrm{in}} = \ket{+}\!\bra{+}$. The adversary ideally optimises the initial state $\rho_B^{\mathrm{in}}$ to capture the most amount of information. Let us define conditional spread $\Gamma$ between the channels as
\begin{align}
  \Gamma \defeq \max_{x,x'} \frac{1}{2}\dnorm{\Chan_B^{x} - \Chan_B^{x'}}.
  \label{eq:eps}
\end{align}
Conditional spread $\Gamma$ measures the largest amount by which the secret $x$ impacts the adversary's observations by quantifying how far apart the two most different conditional channels sit from each other. This is different from the average gate infidelity, $r$, a device datasheet reports, which measures the average per-gate error probability. This paper makes a case for reporting $\Gamma$ to the public for security and privacy analysis prior to using the cloud services. This would be of utmost importance to tenants who run quantum circuits on private sensitive data, e.g., in defence or healthcare industries.

We adopt a notion of information leakage based on distinguishability of the quantum states $\{\rho_B^x\}_{x\in \Xset}$ to measure the adversary's performance in terms of guessing secret $x$ gets once it looks at $B$. Information leakage is defined as
\begin{equation}
  \Leak(A \! \to \! B) \defeq
  \log_2 \sup_{\{\Pi_x\}} \sum_{x \in \Xset} \Tr( \Pi_x \rho_B^{x} ),
  \label{eq:leakdef}
\end{equation}
where the supremum runs over all measurements $\{\Pi_x\}$ the adversary could perform, indexed by the secret it is guessing. The notion of information leakage~\eqref{eq:leakdef} is the logarithm of $M$ times the maximum success probability of distinguishing the quantum states $\{\rho_B^x\}_{x\in \Xset}$ under uniform prior~\cite{barnett2009quantum}. Information leakage $\Leak(A \! \to \! B)$ is measured in bits and sits between zero (no information gained) and $\log_2M$ (maximum amount of information that can be gained about an $M$-ary random variable). This is exactly the maximal quantum leakage of~\cite{ref:farokhi_leakage}, originally defined there as the largest factor by which observing $B$ improves the adversary's chance of guessing any function of secret $x$, and shown to be equal to the Sibson mutual information of order infinity between the victim's secret and the adversary's measurement outcome. We use the state-discrimination form above because it is what the rest of this paper computes directly. The equivalence between the two forms is proved in~\cite{ref:farokhi_channel}. This notion of information leakage needs no assumption about how likely each secret is. It is zero exactly when every $\rho_B^x$ is the same state since, in that case, $B$ carries no information about $x$ at all. Furthermore, via the connection to maximal leakage, we can see that this notion does not make any assumptions about the adversary's intended goal, i.e., what aspect of the secret $x$ the adversary is interested in learning. In summary, a leakage of zero bits means co-tenancy does not help the adversary in learning the secret at all. A leakage of one bit means looking can reveal the answer to a single yes-or-no question about the secret perfectly. \ref{sec:leak} describes maximal leakage and its connection to distinguishability and~\eqref{eq:leakdef} concretely.

\subsection*{Leakage is controlled by conditional spread.}

\ref{proof:result_1} proves the following inequality relating information leakage and conditional spread.

\emph{Result 1.} For any boundary interaction and any victim state, 
\begin{align}
  \Leak(A \! \to \! B) &\le \min\{ \log_2 M,\ \log_2( 1 + (M-1)\Gamma ) \}.
  \label{eq:result1}
\end{align}
For binary secret $\Xset=\{0,1\}$, this can be tightened to
\begin{align}
\Leak(A \! \to \! B) =\log_2 \left( 1 + \frac{1}{2}\|\rho_B^{(0)}-\rho_B^{(1)} \|_1 \right)
  \label{eq:result0}
\end{align}
using the Helstrom bound~\cite{Helstrom1969,Helstrom1976}. If $M\Gamma\ll 1$, which should be the case in high-performance cloud computers, $\Leak(A \! \to \! B) \simeq (M-1)\Gamma/\ln2$ implying that a single run of the victim's circuit does not leak much information about the secret. However, given circuits are run several times due to hardware noise and stochastic nature of most quantum computing algorithms, the leakage can add up to a significant amount. 

\begin{figure*}[t]
\includegraphics[width=\textwidth]{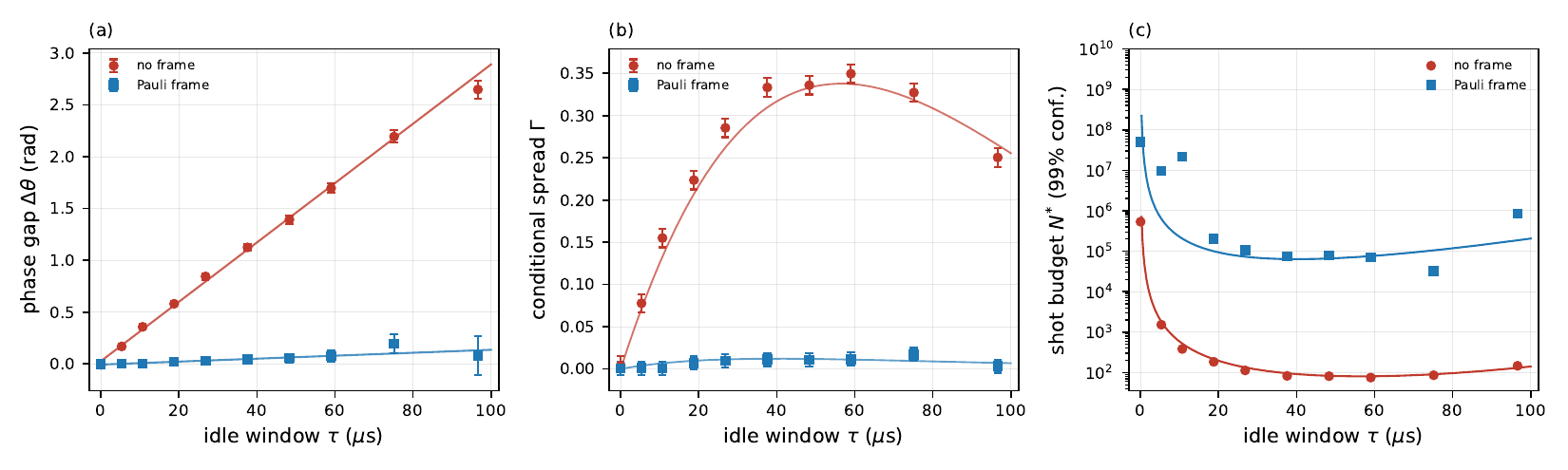}
\caption{Cross-tenant information leakage measured on \texttt{ibm\_kingston}, victim $q_{60}$ and adversary $q_{61}$ using $4096$ shots per circuit with $4$ circuits used for the experiment. (a) Conditional phase gap $\Delta \theta$ against idle window length $\tau$. (b) Conditional spread $\Gamma$ against idle window length $\tau$. (c) Adversary shot budget $N^*$ at $99\%$ confidence against idle window length $\tau$. The markers indicate experimental data. The solid line shows the theoretical relationship fitted to the experimental data. The red curve shows the results without Pauli frame defence while the blue curve shows the outcome with Pauli frame defence.}
\label{fig:data}
\end{figure*}

\subsection*{Repetition can amplify small per-shot leakage.}

In quantum computing, a circuit is often run more than once. Therefore, the adversary can hold up to $N$ copies of its conditional state. For distinguishability of quantum states, standard large-deviations theory gives an error probability $P_e^{(N)}\doteq e^{-N\xi}$~\cite{ref:qcb}, where $\doteq$ means equality of the exponential rate, $\lim_{N\to\infty} -N^{-1}\ln P_e^{(N)} = \xi$. 

Two different values of $\xi$, corresponding to two different scenarios, matter here. An adversary who can measure all $N$ copies jointly, which is only possible with access to stable quantum memory to store the states before a collective measurement, achieves $\xi_{\mathrm{coll}}$~\cite{ref:qcb}.  However, an adversary with no quantum memory, who must measure and discard each copy, one shot at a time, achieves $\xi_{\mathrm{loc}}=-\ln(1-\Gamma^{2})/2\simeq\Gamma^{2}/2$ for $\Gamma\ll 1$, which is smaller than $\xi_{\mathrm{coll}}$. This follows from the classical Chernoff bound applied to the resulting Bernoulli statistics~\cite{chernoff1952measure} of single-shot measurements. Given the current state of quantum computers and memory (particularly the short decoherence time of qubits), the second, weaker adversary is the realistic one. The weaker adversary's shot budget, i.e., the minimum number of measurements required, for confidence $1-\eta$ is
\begin{equation}
  N^{*} \simeq \frac{2\ln(1/\eta)}{\Gamma^{2}}.
  \label{eq:nstar}
\end{equation}
Note that this is only an approximation as we are considering the regimes of $\Gamma\ll 1$ and $N\gg 1$. The leakage of $N$ copies can also be written directly as
\begin{equation}
\mathcal{L}_N(A\to B)\triangleq\log_2\sup_{\{\Pi_x\}}\sum_{x\in\mathcal{X}}
\operatorname{Tr}\!\left(\Pi_x\,\rho_B^{x\otimes N}\right),
\end{equation}
where the supremum is taken over POVMs on the $N$-copy space, which
permits entangled measurements. For a binary secret, by the Helstrom bound~\cite{Helstrom1969,Helstrom1976},
\[
\mathcal{L}_N(A\to B)=\log_2\!\left(1+\frac12\|\rho_B^{0\otimes N}-\rho_B^{1\otimes N}\|_1\right).
\]
By the quantum Chernoff bound~\cite{ref:qcb},  $\lim_{N\to\infty}\mathcal{L}_N(A\to B)=\log_2 M$ if and only if the states $\{\rho_B^{x}\}_{x\in\mathcal{X}}$ are pairwise distinct. Hence, even if a single circuit run leaks little because $\Gamma\ll 1$, repetition weakens any such guarantee.

\subsection*{Average infidelity does not determine leakage.}

Now, we investigate the relationship between information leakage and  average gate infidelity, which is a number that the device datasheet contains. We construct two scenarios with the same average gate infidelity, but with widely different leakage.

First, consider the case where the boundary interaction rotates the probe state of the adversary by angle $\theta$ for one value of the victim's secret (e.g., $x=0$) and by $-\theta$ for the other value (e.g., $x=1$). In this case, $r = (2/3)\sin^2(\theta/2)$ and $\Gamma = |\sin\theta|$. Therefore, for small $\theta$, $\Gamma\approx |\theta|\approx \sqrt{6r}$. This is similar to the case we consider in our experiments.  Eq.~\eqref{eq:nstar}, in this case, gives the shot budget for confidence $1-\eta$ at fixed $r$,
\begin{equation}\label{eq:coh}
  N^{*}_{\mathrm{coh}} \simeq \frac{\ln(1/\eta)}{3r}.
\end{equation}
As an alternative, consider the case where the boundary interaction perturbs the probe state of the adversary by a Pauli channel with a single-qubit $Z$-flip of probability $p$ regardless of victim's secret value. In this case, $r=(2/3)p$ and $\Gamma=0$. Therefore, $\mathcal{L}(A\to B)=0$. No number of shots reveals
anything about the secret, and the shot budget is infinite for every
value of $r$. In contrast, the coherent boundary with the same
average gate infidelity $r$ has the finite shot budget in
Eq.~\eqref{eq:coh}.

This comparison demonstrates the paper's central message. Two
boundaries whose benchmarked average gate infidelities agree may
result in vastly different information leakage, from zero to a
finite shot budget that scales as $r^{-1}$. This is because average gate infidelity cannot see the coherent-versus-stochastic split while conditional spread can. A privacy-relevant certificate therefore needs a measure of coherence in addition to error. 

\section*{Defence against leakage}

If leakage tracks coherence and fidelity tracks total error, a protocol that removes coherence without changing error should remove leakage without changing fidelity. This builds on the Pauli twirling used in randomised compiling, which converts coherent errors into stochastic Pauli noise and has already been demonstrated experimentally on hardware~\cite{ref:wallman_rc, ref:hashim,ref:crosstalk_rc}. Therefore, the machinery this defence needs is not new. Assume that, in every run of the circuit, the victim draws a uniform random Pauli operator $P=\bigotimes_{a\in A}P_a$ over its \emph{entire} register, applies it with the matching correction after, and never reveals which $P$ it drew. We call this random, secret, self-correcting operation a \emph{Pauli frame}. A deployment tuned only for error mitigation need not keep the draw secret. However, for the adversary to not be able to infer the secret, we need to keep the draw secret. The next result, proved in~\ref{proof:result_2}, shows that the leakage is zero with Pauli frame defence.  

\emph{Result 2.} Assume that the victim's input state $\rho_A^{(\mathrm{in})x}$ depends on secret $x$ (so the Hamiltonian is independent of $x$). Under the Pauli frame described above, $\Gamma=0$ and $\Leak(A\to B)=0$. 

This is a quantum one-time pad on the victim's register~\cite{boykin2003optimal}. The proof needs no assumption on the coupling because, once the frame has erased the victim's state, there is nothing left in $A$ for the coupling to carry across. The frame itself is the pad's key. Therefore, the secrecy of the draw during the observation window is a pivotal assumption. The defence protects a tenant against a co-tenant, not against a provider that can read the submitted circuit directly. A cheaper version of the Pauli frame that only implements the random Pauli operator on boundary qubits $\partial A$, rather than the whole register $A$, does {not} work. A randomisation confined to $\partial A$ only perturbs the boundary qubits at the start of the circuit run. However, during the operation of victim's own circuit, the content of some of the non-boundary qubits can move onto the boundary and leak private information. Result~2 assumes that the Hamiltonian does not depend on the secret. However, the couplings $\zeta_k^{(x)}(t)$ can themselves depend on the secret $x$. The following result provides when the leakage can be zero in this case. The following result immediately follows from Result~1.

\emph{Result 3.} $\Leak(A\!\to\! B)=0$ if $\Chan_B^{x}=\Chan_B^{x'}$ for every pair $x,x'$, i.e., the
adversary's conditional channel remains the same regardless of the secret.

This is the criterion a provider should actually certify. Result~2 gives one way to guarantee it for scenarios in which the secret is only embedded in the victim's state, and not the Hamiltonian.

\begin{figure}[t]
    \centering
    \includegraphics[width=1\linewidth]{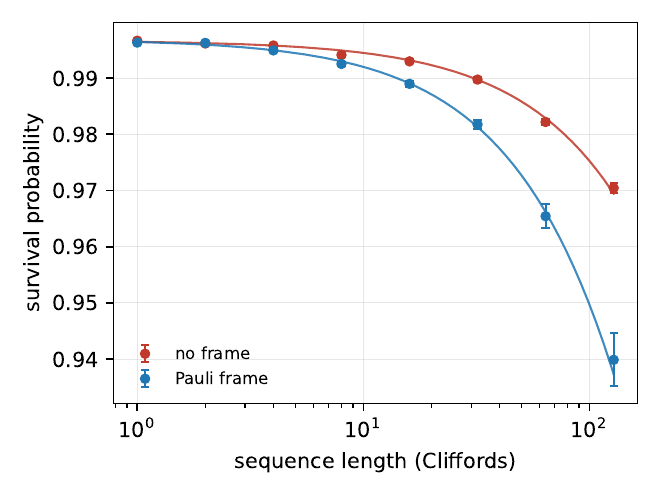}
    \caption{Randomised benchmarking on victim qubit $q_{60}$ on \texttt{ibm\_kingston}. Survival probability against the number of random single-qubit Clifford gates chained together before the sequence-ending inverse gate demonstrates that the Pauli frame defence non-trivially degrades the victim circuit.}
    \label{fig:fidenlity}
\end{figure}

\begin{figure*}[t]
\includegraphics[width=\textwidth]{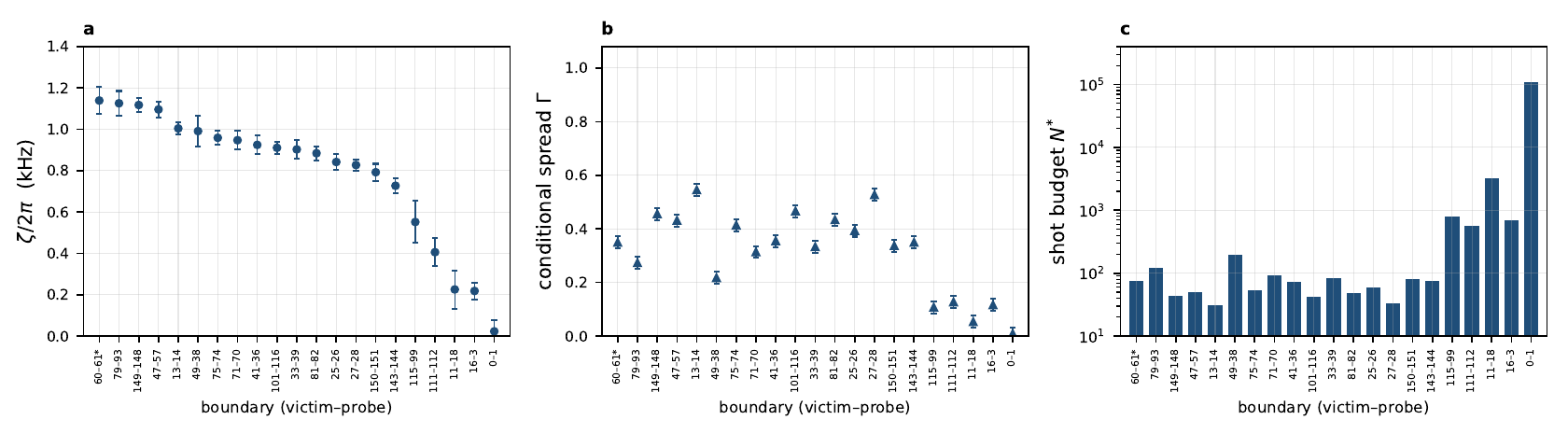}
\caption{Twenty one boundaries surveyed on \texttt{ibm\_kingston} using $1000$ shots per circuit with 4 circuits used for the experiment. (a) Coupling rate $\zeta/2\pi$ per boundary. (b) Conditional spread $\Gamma$ per boundary. (c) Adversary shot budget $N^{*}$ at $99\%$ confidence  per boundary. }
\label{fig:survey}
\end{figure*}

\subsection*{Experimental measurement of the attack.}
We demonstrate the potential for information leakage on \texttt{ibm\_kingston}, a $156$-qubit superconducting noisy intermediate-scale quantum (NISQ) processor (IBM Heron r2). We subsequently verify the information leakage on \texttt{iqm garnet}, a $20$-qubit superconducting NISQ processor (IQM). 

The interaction that we realise on the hardware experiment is summarised in Figure~\ref{fig:threat}. The victim holds qubit $a$ on its sector $A$, which is prepared in $\ket{0}$ or $\ket{1}$. The choice of this preparation is the secret. The adversary holds an adjacent qubit $b$ on it sector $B$. It prepares $\ket{+}$ on qubit $b$, idles for a window $\tau$, and reads out a measurement in $X$ and $Y$ basis.  The victim's qubit is never measured by the adversary. The victim's boundary qubit $a$ and an adversary's probe qubit $b$ are coupled by a static $ZZ$ interaction. The rate of the interaction is defined as $\zeta$. The two conditional probe states, i.e., the probe state when the victim holds $\ket{0}$ and the probe state when the victim holds $\ket{1}$,  are separated by phase gap $\Delta\theta = 4\zeta\tau$. Therefore, 
\begin{align}\label{eqn:gamma_T1_T2}
    \Gamma = C(\tau)\,|\sin(\Delta\theta/2)|
\end{align}
where $C(\tau)=\exp(-\tau/T_2)$ is a number between zero and one that falls as the probe loses coherence with increasing $\tau$. The reason for preparing $\ket{+}$ and the derivation of~\eqref{eqn:gamma_T1_T2} are discussed in~\ref{app:optimality_of_plus}. The procedure for experimentally measuring $\Delta \theta$ and $\Gamma$ is described in \ref{app:experiment}. 

Figure~\ref{fig:data}(a) shows that the experimentally estimated $\Delta\theta$ grows linearly in $\tau$, as expected.  Figure~\ref{fig:data}(b) shows
$\Gamma$, which peaks at $0.35$ near $60\mu$s. This gives a fitted value for $T_2$ equal to $73.3\mu$s, which is of the order of the device's reported coherence time $T_2 = 53.7\mu$s. At the peak the adversary extracts $\Leak (A\to B)= 0.43$ bits per shot, out of one possible bit $\log_2 M=1$. Figure~\ref{fig:data}(c) compares the predicted shot budget of Eq.~\eqref{eq:nstar} based on the hardware test.  
The markers indicate the experimental data. The solid curve shows the theoretical relationship fitted to the experimental data. In all figures, the blue curve shows the effect of a randomising defence defined formally as the Pauli
frame in Result~2. The defence clearly works by massively reducing information leakage and increasing the number of shots needed to estimate the victim's private data confidently.

\emph{The suppression is paid for in fidelity.} To test whether the Pauli frame defence in Result~2 costs the victim anything on its own qubit, we performed randomised benchmarking on $q_{60}$, the same victim qubit used earlier to show the success of the Pauli frame defence. A sequence of length $m$ consists of $m$ Cliffords drawn uniformly at random from the $24$ single-qubit Cliffords, followed by one further gate computed deterministically as the exact group inverse of their combined effect, so that, under no error, a circuit initialised at $\ket{0}$ returns exactly to $\ket{0}$. We run two alternative scenarios of applying the gates without any defence and with the Pauli frame defence. For the latter, a random Pauli operator is applied before each gate and corrected for after every gate in the sequence. Both runs accumulate the same net Clifford and differ only in the Pauli frame's overhead. Survival probability is the fraction of the runs that initialised at $\ket{0}$ returned to $\ket{0}$. Figure~\ref{fig:fidenlity} shows the survival probability as a function of number of randomly drawn Cliffords. The defence impacts the survival probability by a non-trivial amount. That is, although theoretically suppressing information leakage does not degrade the victim's circuit, in practice, there is an impact due to accumulation of error with application of each extra gate.  

\emph{The effect is not one lucky pair.} We repeated the same experiment on twenty-one distinct boundaries. Motivated by the earlier observation that $\Gamma$ is peaked near $T_2$, we set $\tau$ for each pair based on its reported $T_2$. In fact, 
we set the idle window of each boundary as the smaller of the reported probe $T_2$ and one third of the reported victim $T_1$. The window is capped at one third of the victim $T_1$ so that the victim's basis state does not relax appreciably during the idle window. Figure~\ref{fig:survey}(a) shows the estimated coupling rate $\zeta/2\pi$. The coupling rate has median
$903$~Hz, with a 25th-to-75th-percentile range of $727$ to $992$~Hz. The coupling rate exceeds $250$~Hz on $18$ boundaries and $750$~Hz on $15$. The pair 60-61, studied earlier, sits at the upper end of this spread, but it is not atypical. Figure~\ref{fig:survey}(b) shows conditional spread $\Gamma$ for all the pairs. The conditional spread has a median $0.35$. Figure~\ref{fig:survey}(c) shows the shot budget $N^*$ at 99\% confidence, calculated based on the estimated conditional spread. The adversary's shot budget $N^*$ has median $75$. 

\begin{figure*}[t]
\centering
\includegraphics[width=\textwidth]{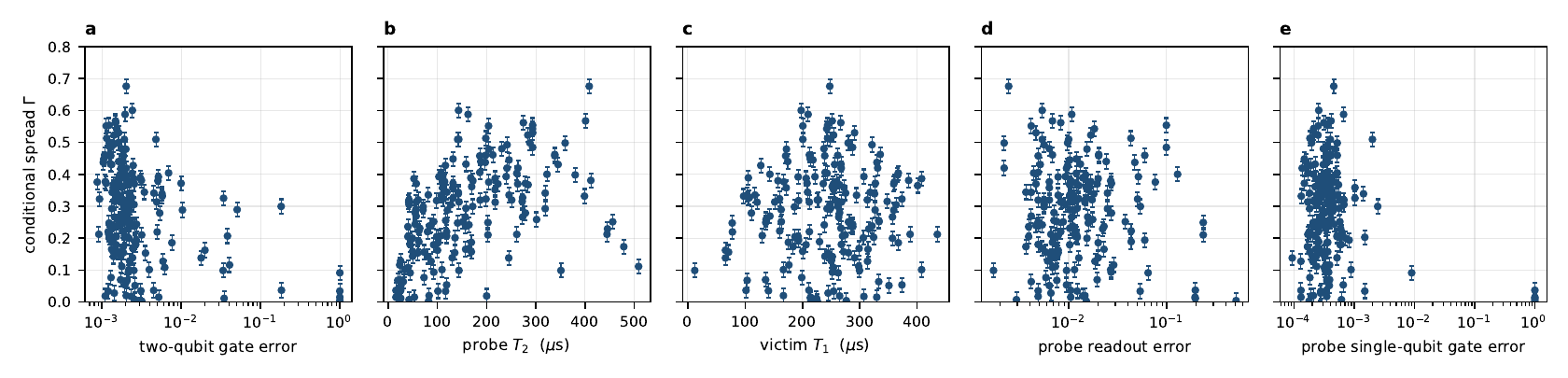}
\caption{Measured conditional spread $\Gamma$ against the numbers reported by the device, on every coupled pair of \texttt{ibm\_kingston}, using 1000 shots per circuit and 4 circuits per boundary. Each point is one of the 176 boundaries. (a) Reported two-qubit gate error on a logarithmic axis. (b) Reported probe $T_2$. (c) Reported victim $T_1$. (d) Reported probe readout error on a logarithmic axis. (e) Reported probe single-qubit gate error on a logarithmic axis.}
\label{fig:datasheetcomparison}
\end{figure*}

\textit{Reported device numbers are not enough.} We repeated the boundary survey on every coupled pair of \texttt{ibm\_kingston}, 176 boundaries in total, using the same four circuits per boundary and 1000 shots. Similarly, the idle window of each boundary was the smaller of the reported probe $T_2$ and one third of the reported victim $T_1$. Figure~\ref{fig:datasheetcomparison} shows $\Gamma$ against the reported two-qubit gate error, the probe $T_2$, the victim $T_1$, the probe readout error and the probe single-qubit gate error. Of the 176 boundaries, 134 have a reported two-qubit gate error between $1\times10^{-3}$ and $3\times10^{-3}$, and across these $\Gamma$ ranges from 0.007 to 0.68. The boundaries 20--21 and 137--147 have the same reported two-qubit gate error of $2.03\times10^{-3}$ and measured $\Gamma$ of 0.68 and 0.20. More generally, Figure~\ref{fig:datasheetcomparison} shows that, for each of the five reported quantities from the device, boundaries with similar values of that quantity cover a wide range of $\Gamma$. The reported two-qubit gate error shows a tendency toward lower $\Gamma$ at the largest reported errors, with wide scatter at all values. The conditional spread versus the probe $T_2$ shows a general upward trend with large scatter around it. This trend is not independent of the choice of idle window $\tau$, since the window is set partly from the probe $T_2$. The victim $T_1$, the probe readout error, and the probe single-qubit gate error show no clear trend. Therefore, on this device, the reported numbers alone are not enough to determine the leakage of a given boundary. Reporting information leakage, or a proxy such as the conditional spread, is of utmost importance to enable understanding the privacy and security risks associated with co-tenancy.

\emph{Boundary randomisation is not enough.} Consider the case where the victim holds an interior qubit $q_{150}$ while $q_{149}$ is the boundary. The victim's secret is held on the interior qubit $q_{150}$ and the secret encodes whether the state is prepared at $\ket{0}$ or $\ket{1}$. The adversary's probe is on $q_{148}$ and is initialised at $\ket{+}$. These qubits are in a line. After the initialisation, a defence mechanism is implemented. Then, a SWAP that moves the secret from $q_{150}$ onto $q_{149}$, which, in this experiment, models the victim's legitimate circuit operation, then the idle window $\tau$, where the $q_{149}$ and $q_{148}$ get coupled through unwanted crosstalk, then any correction due to the defence. After this, the adversary reads $q_{148}$. We implement two defence mechanisms. The first mechanism is the one described in Result~2, which involves random Pauli operations on both qubits $q_{149}$ and $q_{150}$. This should theoretically render the adversary completely incapable of estimating the secret held by the victim. The second approach is the heuristic of implementing random Pauli operations on the boundary qubit $q_{149}$. We compare these with the baseline of implementing no defence. 

\begin{figure*}[t]
\includegraphics[width=\textwidth]{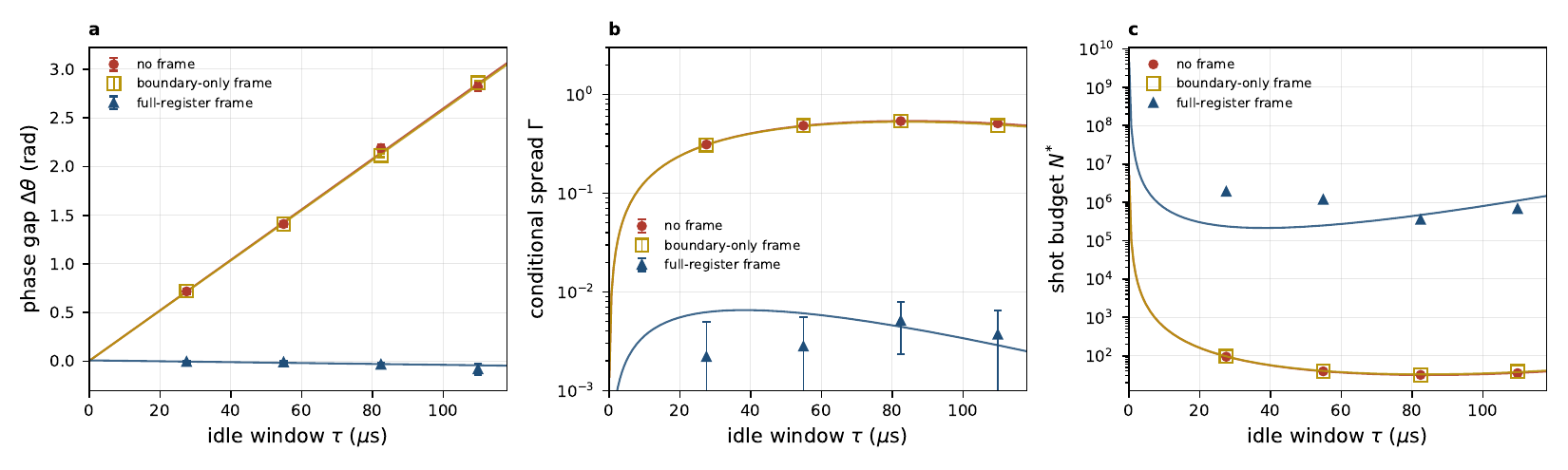}
\caption{Effect of imperfect defence by randomisation on boundary qubits, where the victim circuit moves the secret from an interior qubit onto the boundary qubit measured on \texttt{ibm\_kingston}. (a) Conditional phase gap against idle window length. (b) Conditional spread $\Gamma$ against idle window length. (c) Adversary shot budget at 99\% confidence. The heuristic boundary-only frame sits on top of the undefended baseline, i.e.,  it does not work as a defence mechanism. The Pauli frame from Result~2 suppresses the leakage significantly.
}
\label{fig:hierarchy}
\end{figure*}

Figure~\ref{fig:hierarchy} shows the conditional phase gap $\Delta \theta$, conditional spread $\Gamma$, and the adversary shot budget at 99\% confidence against idle window length $\tau$. The cheaper defence achieves nothing. However, the leakage is considerably suppressed under the full frame defence of Result~2. The boundary-only frame recovers Result~2 only in the special case where the part of victim's state containing secret $x$ never reaches $\partial A$ during the window. For a victim running an actual circuit with potential movement of the secret towards the boundary qubits, a boundary-only frame gives an illusion of protection. Only the full-register frame of Result~2 closes the channel.

\emph{The residual, and where Result 2 meets hardware.}
Result~2 predicts exactly zero leakage under the full-register frame. What we measure is a small residual leakage, which is small enough to practically imply no detection as it pushes the shot counts to millions. The residual leakage can be due to noisy observations with finite number of shots. In addition, it may be also caused by an assumption that the hardware violates. A transmon,
the superconducting circuit used here, is not a clean two-level qubit, and admits a third energy level. A Pauli frame built from operators on the two-level subspace does not symmetrise the third level. Result~2 is exact in the ideal qubit model, but on real hardware it potentially carries a correction set by how much information is leaked from that third level. Separating the impact of these issues is an interesting avenue for future research.

\begin{figure}[t]
    \centering
    \includegraphics[width=1\linewidth]{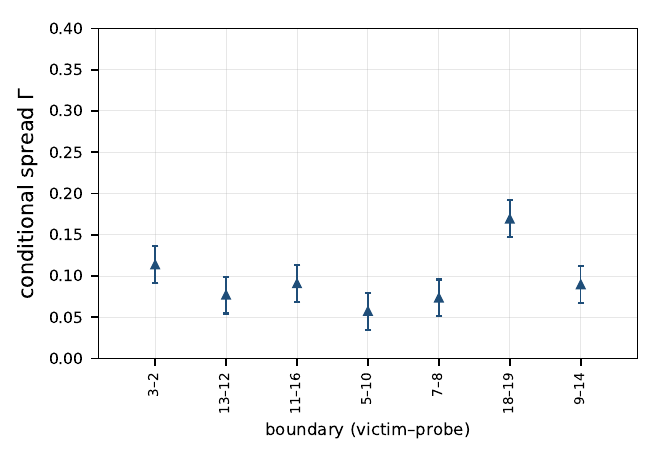}
    \caption{Conditional spread $\Gamma$ on seven boundaries surveyed on \texttt{iqm garnet} using $1000$ shots per circuit with 4 circuits used for the experiment. }
    \label{fig:iqm}
\end{figure}

\emph{The observation is not vendor specific.} To check whether this result is general, we repeated a reduced version of the survey on a different vendor's device, \texttt{iqm garnet}, accessed through IQM's Resonance cloud service. Seven potential tenant boundaries were
measured with the idle window chosen per boundary from each qubit's own measured coherence time. Figure~\ref{fig:iqm} illustrates the conditional spread across the boundaries. The conditional spreads are smaller than those measured on \texttt{ibm\_kingston}, but this is expected due to the devices substantially shorter $T_2$, which has a median of $10.3\mu$s against several tens of microseconds on \texttt{ibm\_kingston}. This forces a correspondingly shorter idle window $\tau$ and therefore less accumulated phase before decoherence. 

\section*{Discussion}
The privacy of a shared quantum processor is set by the coherence of its tenant boundary, not by the boundary's error rate. The figure of merit providers currently publish cannot support a privacy guarantee on its own. Repetition weakens any guarantees provable per circuit run. 
Victim-side Pauli randomisation over the whole register results in a quantum one-time pad and closes this channel. The cheaper boundary-only version of the same idea provides no protection. Randomised compiling, already used industrially to tailor noise, provides a promising protection. Two issues require further investigation. We surveyed boundaries on two  superconducting quantum computers. Showing this issue experimentally on other types of devices remains open. Also, scaling the bounds here to many co-scheduled tenants at once (not just one victim and one adversary) raises a composition question much like the one differential privacy similarly faces~\cite{alabi2026does}.

\section*{Methods}
An important choice, relating to  control settings that must be turned off, matters here for interpreting the results. Particularly, gate and measurement twirling must be disabled. Gate twirling is precisely the Pauli frame defence under test. Leaving it on erases the  attack and would be mistaken for a null result. Dynamical decoupling must also be disabled, because it actively echoes away the static $ZZ$ interaction being measured. Note that, given these can be turned off by the victim and the adversary, it is important to document their importance and relevance to information leakage. Analysis code, circuit generators, and the data behind every figure are available at~\cite{code}.

\subsection*{Use of large language models}
During preparation of this manuscript the author used Claude Sonnet 5 (Anthropic), a large language model, for revising the prose and  writing the Python and Qiskit code used for hardware experiments. All AI outputs were reviewed, independently verified against exact numerical computation or primary literature where applicable, and edited by the author. The author takes full responsibility for the  content of the published article. 

\section*{Data availability}
The processed data underlying every figure is available at~\cite{code}. Backend calibration data for \texttt{ibm\_kingston} and \texttt{iqm garnet} are available through their respective cloud provider's platform~\cite{ibmqpuinfo, iqmqiskitguide}.

\section*{Code availability}
All circuit-generation, execution, and analysis code is available at~\cite{code}.

\section*{Competing interests}
This work used IBM Quantum computing access provided through the IBM Quantum Network Hub at the University of Melbourne. The author declares no competing financial interests. 

\section*{Ethical and dual-use considerations}
The underlying security threat, information leakage or interference through crosstalk in shared/multi-tenant quantum processors, has been publicly demonstrated, including on IBM quantum hardware~\cite{ref:choudhury,ref:harper,ref:mehra}. Our contribution is the quantitative information-theoretic characterisation of this leakage, its relationship to gate infidelity/coherence, and the demonstrated mitigation. We judge the net effect of publication to be protective rather than harmful, since the paper's primary contribution is the defence and the call-for-action for transparent data release from the cloud service providers that follows from it, and withholding the quantitative characterisation would leave affected users without a way to assess their exposure while leaving the underlying hardware behaviour unchanged. 

\section*{Author contributions}
F.F. designed the study, performed the theoretical analysis, conducted the experiments, and wrote the
manuscript.

\section*{Acknowledgements}
This work was supported by the University of Melbourne through the establishment of an IBM Quantum Network Hub at the University.

\bibliography{references}

\begin{thebibliography}{10}

\bibitem{ref:multiprog}
P.~Das, S.~S. Tannu, P.~J. Nair, and M.~Qureshi, ``A case for multi-programming
  quantum computers,'' in {\em Proceedings of the 52nd Annual IEEE/ACM
  International Symposium on Microarchitecture (MICRO-52)}, pp.~291--303, 2019.

\bibitem{ref:sarovar}
M.~Sarovar, T.~Proctor, K.~Rudinger, K.~Young, E.~Nielsen, and R.~Blume-Kohout,
  ``Detecting crosstalk errors in quantum information processors,'' {\em
  Quantum}, vol.~4, p.~321, 2020.

\bibitem{ref:zhou}
Z.~Zhou, R.~Sitler, Y.~Oda, K.~Schultz, and G.~Quiroz, ``Quantum crosstalk
  robust quantum control,'' {\em Physical Review Letters}, vol.~131, p.~210802,
  2023.

\bibitem{ref:choudhury}
N.~Choudhury, C.~N. Mude, S.~Das, P.~C. Tikkireddi, S.~Tannu, and K.~Basu,
  ``Crosstalk-induced side channel threats in multi-tenant {NISQ} computers,''
  {\em arXiv preprint arXiv:2412.10507}, 2024.

\bibitem{ref:swapattack}
W.~J.~B. Lee, S.~Wang, S.~Dutta, W.~E. Maouaki, and A.~Chattopadhyay, ``{SWAP}
  attack: Stealthy side-channel attack on multi-tenant quantum cloud system,''
  {\em arXiv preprint arXiv:2502.10115}, 2025.

\bibitem{1011453708821}
W.~J.~B. Lee, S.~Wang, S.~Dutta, W.~E. Maouaki, and A.~Chattopadhyay, ``Poster:
  Stealthy {SWAP}-based side-channel attack on multi-tenant quantum cloud
  systems,'' in {\em Proceedings of the 20th ACM Asia Conference on Computer
  and Communications Security}, ASIA CCS '25, (New York, NY, USA),
  p.~1788–1790, 2025.

\bibitem{ref:readout}
S.~Maurya, C.~N. Mude, B.~Lienhard, and S.~Tannu, ``Understanding side-channel
  vulnerabilities in superconducting qubit readout architectures,'' in {\em
  2024 IEEE International Conference on Quantum Computing and Engineering
  (QCE)}, vol.~1, pp.~1177--1183, IEEE, 2024.

\bibitem{ref:mehra}
D.~Mehra and A.~Kalev, ``Towards defending crosstalk-mediated attacks in
  multi-tenant quantum computing,'' {\em Physica Scripta}, vol.~101, no.~9,
  p.~095102, 2026.

\bibitem{ref:harper}
B.~Harper, B.~Tonekaboni, B.~Goldozian, M.~Sevior, and M.~Usman, ``Crosstalk
  attacks and defence in a shared quantum computing environment,'' {\em
  Advanced Quantum Technologies}, vol.~8, p.~e2500009, 2025.

\bibitem{ref:bell}
B.~Bell, A.~Tr{\"u}gler, K.~Beyer, and P.~Erker, ``Hardware-agnostic modeling
  of quantum side-channel leakage via conditional dynamics and learning from
  full correlation data,'' {\em arXiv preprint arXiv:2602.15966}, 2026.

\bibitem{barnett2009quantum}
S.~M. Barnett and S.~Croke, ``Quantum state discrimination,'' {\em Advances in
  Optics and Photonics}, vol.~1, no.~2, pp.~238--278, 2009.

\bibitem{ref:farokhi_leakage}
F.~Farokhi, ``Maximal information leakage from quantum encoding of classical
  data,'' {\em Physical Review A}, vol.~109, p.~022608, 2024.

\bibitem{ref:farokhi_channel}
S.~Xiao, Z.~Zhao, J.~Zhu, and F.~Farokhi, ``Maximal quantum leakage:
  operational interpretation and quantum channel analysis,'' {\em arXiv
  preprint arXiv:2607.15853}, 2026.

\bibitem{Helstrom1969}
C.~W. Helstrom, ``Quantum detection and estimation theory,'' {\em Journal of
  Statistical Physics}, vol.~1, no.~2, pp.~231--252, 1969.

\bibitem{Helstrom1976}
C.~W. Helstrom, {\em Quantum Detection and Estimation Theory}, vol.~123 of {\em
  Mathematics in Science and Engineering}.
\newblock New York: Academic Press, 1976.

\bibitem{ref:qcb}
K.~M.~R. Audenaert, J.~Calsamiglia, R.~Mu{\~n}oz-Tapia, E.~Bagan, L.~Masanes,
  A.~Ac{\'i}n, and F.~Verstraete, ``Discriminating states: The quantum chernoff
  bound,'' {\em Physical Review Letters}, vol.~98, p.~160501, 2007.

\bibitem{chernoff1952measure}
H.~Chernoff, ``A measure of asymptotic efficiency for tests of a hypothesis
  based on the sum of observations,'' {\em The Annals of Mathematical
  Statistics}, pp.~493--507, 1952.

\bibitem{ref:wallman_rc}
J.~J. Wallman and J.~Emerson, ``Noise tailoring for scalable quantum
  computation via randomized compiling,'' {\em Physical Review A}, vol.~94,
  p.~052325, 2016.

\bibitem{ref:hashim}
A.~Hashim, R.~K. Naik, A.~Morvan, J.-L. Ville, B.~Mitchell, J.~M. Kreikebaum,
  M.~Davis, E.~Smith, C.~Iancu, K.~P. O'Brien, I.~Hincks, J.~J. Wallman,
  J.~Emerson, and I.~Siddiqi, ``Randomized compiling for scalable quantum
  computing on a noisy superconducting quantum processor,'' {\em Physical
  Review X}, vol.~11, p.~041039, 2021.

\bibitem{ref:crosstalk_rc}
H.~Perrin, T.~Scoquart, A.~Shnirman, J.~Schmalian, and K.~Snizhko, ``Mitigating
  crosstalk errors by randomized compiling: simulation of the {BCS} model on a
  superconducting quantum computer,'' {\em Physical Review Research}, vol.~6,
  p.~013142, 2024.

\bibitem{boykin2003optimal}
P.~O. Boykin and V.~Roychowdhury, ``Optimal encryption of quantum bits,'' {\em
  Physical Review A}, vol.~67, no.~4, p.~042317, 2003.

\bibitem{alabi2026does}
D.~Alabi and T.~Nuradha, ``When does quantum differential privacy compose?,''
  {\em arXiv preprint arXiv:2601.00337}, 2026.

\bibitem{code}
F.~Farokhi, ``Code and data for ``{C}oherence rather than error rate governs
  privacy in multi-tenant quantum computing''.''
  \url{https://github.com/farhadfarokhigit/multi-tenant_quantum_leakage}, 2026.

\bibitem{ibmqpuinfo}
{IBM Quantum}, ``View backend details.'' IBM Quantum Platform documentation,
  2026.
\newblock Accessed 28 September 2026.

\bibitem{iqmqiskitguide}
{IQM}, ``Qiskit on {IQM} user guide.'' IQM client documentation, 2026.
\newblock Accessed 28 September 2026.

\bibitem{ref:issa}
I.~Issa, A.~B. Wagner, and S.~Kamath, ``An operational approach to information
  leakage,'' {\em IEEE Transactions on Information Theory}, vol.~66, no.~3,
  pp.~1625--1657, 2020.

\bibitem{krantz2019quantum}
P.~Krantz, M.~Kjaergaard, F.~Yan, T.~P. Orlando, S.~Gustavsson, and W.~D.
  Oliver, ``A quantum engineer's guide to superconducting qubits,'' {\em
  Applied Physics Reviews}, vol.~6, no.~2, p.~021318, 2019.

\bibitem{watrous2018theory}
J.~Watrous, {\em The Theory of Quantum Information}.
\newblock Cambridge University Press, 2018.

\bibitem{kliesch2021theory}
M.~Kliesch and I.~Roth, ``Theory of quantum system certification,'' {\em PRX
  Quantum}, vol.~2, p.~010201, 2021.

\end{thebibliography}
\bibliographystyle{ieeetr}

\clearpage
\onecolumn
\begin{center}
\textbf{\Large Supplementary Information}\\[4pt]
\textit{Coherence rather than error rate governs privacy in shared quantum
processors}
\end{center}
\vspace{8pt}

\setcounter{section}{0}
\setcounter{equation}{0}
\setcounter{figure}{0}
\setcounter{table}{0}
\renewcommand{\thesection}{Supplementary Note \arabic{section}}
\renewcommand{\thesubsection}{Supplementary Note \arabic{section}.\arabic{subsection}}
\renewcommand{\theequation}{S\arabic{equation}}
\renewcommand{\thefigure}{S\arabic{figure}}
\renewcommand{\thetable}{S\arabic{table}}

\section{Maximal quantum leakage and its normalisation}\label{sec:leak}
The objective of the adversary is estimate or guess a possibly randomised function of the secret $x$, denoted by $z=f(x)$, by performing measurements on states $\{\rho_B^x\}_{x\in\Xset}$. Maximal quantum leakage, defined in~\cite{ref:farokhi_leakage} by extending the classical notion of maximal leakage~\cite{ref:issa}, compares two scenarios. In the first scenario, the adversary performs positive operator-valued measure (POVM) on the states to obtain measurement $y$ and then takes a guess of $z$ denoted by $g(y)$. In the second scenario, the adversary does not measure the states (modelling the case where the adversary's state is independent of $x$ and thus measurement is useless) and takes a guess of $z$ denoted by $\bar{g}$, where $\bar{g}$ is a constant independent of $y$ or $x$. The maximal quantum leakage measures the ratio of the probabilities of success between these two scenarios, that is 
\begin{align} \label{def:qml}
    \mathcal{Q}(A\to B)\defeq\sup_{\text{POVM},f,g,\bar{g}} \log_2 \left(\frac{\mathbb{P}\{z=g(y)\}}{\mathbb{P}\{z=\bar{g}\}} \right),
\end{align}
where the supremum is taken over all targets of attack (determined by $f$), estimation algorithms with access and without access to data (determined by $g$ and $\bar{g}$), all measurement policies (determined by the POVM). The maximal quantum leakage, as characterised in~\eqref{def:qml}, captures the multiplicative \emph{increase} in the probability of correctly guessing any general random or deterministic function of secret $x$ upon accessing the quantum encoding of the data $\{\rho_B^x\}_{x\in\Xset}$. The maximal quantum leakage is proved to be equal to the Sibson mutual information of order infinity between secret $x$ and the adversary’s measurement outcome $y$~\cite{ref:farokhi_leakage}. This notion of leakage is intimately related to distinguishability of the states $\{\rho_B^x\}_{x\in\Xset}$ under uniform prior~\cite{ref:farokhi_channel}:
\begin{align}
    \mathcal{Q}(A\to B)&=\Leak(A\!\to\! B) \\
  &= \log_2 \sup_{\{\Pi_x\}} \sum_{x\in\Xset}\Tr(\Pi_x\rho_B^x),
  \label{eq:s_leak}
\end{align}
where the supremum is taken over POVMs with $M$ outcomes indexed by $\Xset$. Three properties follow from~\cite{ref:farokhi_leakage, ref:farokhi_channel}. Information leakage $\Leak(A\!\to\! B)$ vanishes if and only if $\rho_B^x=\rho_B^{x'}$ for all $x\neq x'$. Information leakage is bounded above by $\log_2 M$. And, for $M=2$ and $\Xset=\{x,x'\}$, the leakage is exactly equal to 
\begin{align}
   \Leak(A\!\to\! B)= \log_2( 1 + \frac{1}{2}\|\rho_B^x - \rho_B^{x'}\|_1)
\end{align}
due to the Helstrom bound~\cite{Helstrom1969,Helstrom1976}.

\section{Proof of Result 1} \label{proof:result_1}
Define the average conditional state as
\begin{align}
    \bar\rho_B = \frac{1}{M}\sum_x \rho_B^x.
\end{align}
Let 
\begin{align}
    \Delta_x = \rho_B^x-\bar\rho_B.
\end{align}
We can see that $\sum_x \Tr(\Pi_x \rho_B^x) = 1 + \sum_x \Tr(\Pi_x\Delta_x)$, and since every measurement operator satisfies $0\le \Pi_x\le I$ while $\Tr\Delta_x=0$,
each term obeys $\Tr(\Pi_x\Delta_x)\le \tnorm{\Delta_x}/2$. Summing over
$x$ and substituting into Eq.~\eqref{eq:leakdef} gives
\begin{equation}
  \Leak(A \! \to \! B) \le
  \log_2\!\left( 1 + \sum_{x} \frac{1}{2}\tnorm{\rho_B^x - \bar\rho_B} \right).
  \label{eq:lemma1}
\end{equation}
Eq.~\eqref{eq:lemma1} holds as an equality, not just a bound, for $M=2$. With only two secrets the sum on the right is exactly the trace distance between the two conditional states and the equality follows from the Helstrom bound~\cite{Helstrom1969,Helstrom1976}. Note that
\begin{equation}
  \rho_B^x-\bar\rho_B=\frac1M\sum_{x'}(\rho_B^x-\rho_B^{x'}).
\end{equation}
By the triangle inequality and $\rho_B^x=\Chan_B^x(\rho_B^{\mathrm{in}})$,
\begin{equation}
  \frac12\tnorm{\rho_B^x-\bar\rho_B}
  \le\frac1M\sum_{x'} \frac12\tnorm{\Chan_B^x(\rho_B^{\mathrm{in}})-\Chan_B^{x'}(\rho_B^{\mathrm{in}})}
  \le\frac1M\sum_{x'}\frac12\dnorm{\Chan_B^x-\Chan_B^{x'}}
  \le\frac{M-1}{M}\Gamma.
\end{equation}
Summing over $x$ gives 
\[
\sum_x\frac12\tnorm{\rho_B^x-\bar\rho_B}\le (M-1)\Gamma,
\]
which, in conjunction with Eq.~\eqref{eq:lemma1}, results in 
\begin{align} \label{eqn:general_bound}
    \Leak(A\to B)\le\log_2(1+(M-1)\Gamma).
\end{align}
Also, note that $\Leak(A\to B)\le\log_2 M$~\cite{ref:farokhi_leakage}. 

\section{Proof of Result 2}\label{proof:result_2}

With the Pauli frame defence, the conditional channel is given by
\begin{align}
    \Chan_B^x(\rho_B^{\mathrm{in}}) =\mathbb{E}_P\left\{
    \Tr_A\left[V_P(\rho_A^{\mathrm{in}(x)}\otimes\rho_B^{\mathrm{in}})V_P^\dag\right]\right\},
\end{align}
where
\begin{align}
    V_P=(C_P\otimes I)\,U(t,\tau)\,(P\otimes I),
\end{align}
where $C_P$ is any correction chosen to keep the victim's intended logical operation correct for the Pauli draw $P$. Here, expectation with respect to the Pauli frame is taken as the adversary does not know the frame $P$ used. For an idle victim $H_A^{(x)}=0$, $C_P=P^\dagger$. 
For the victim implementing any gate $G_A$ (evolution operator under $H_A^{(x)}\neq 0$), the standard choice is $C_P = G_A P^{\dagger} G_A^{\dagger}$. The specific choice of $C_P$ does not matter for what follows. Note that
\begin{align}
    \Chan_B^x(\rho_B^{\mathrm{in}})
    &= \mathbb{E}_P\left\{
    \Tr_A\left[(C_P\otimes I)\,U(t,\tau)\,(P\otimes I)(\rho_A^{\mathrm{in}(x)}\otimes\rho_B^{\mathrm{in}})(P^\dag \otimes I)\,U(t,\tau)^\dag\,(C_P^\dag \otimes I)\right]\right\}\\
    &= \mathbb{E}_P\left\{
    \Tr_A\left[U(t,\tau)\,(P\otimes I)(\rho_A^{\mathrm{in}(x)}\otimes\rho_B^{\mathrm{in}})(P^\dag \otimes I)\,U(t,\tau)^\dag\right]\right\}\\
    &= 
    \Tr_A\left[U(t,\tau)\, \mathbb{E}_P\left\{(P\otimes I)(\rho_A^{\mathrm{in}(x)}\otimes\rho_B^{\mathrm{in}})(P^\dag \otimes I)\right\}\,U(t,\tau)^\dag\right]\\
    &= 
    \Tr_A\left[U(t,\tau)\, (\mathbb{E}_P\{P\rho_A^{\mathrm{in}(x)}P^\dag\}\otimes\rho_B^{\mathrm{in}})\,U(t,\tau)^\dag\right]\\
    &= 
    \Tr_A\left[U(t,\tau)\,(I/2^{n_A}\otimes\rho_B^{\mathrm{in}})\,U(t,\tau)^\dag\right]
\end{align}
where the second equality stems from the properties of trace $\Tr_A[(C_P\otimes I)M(C_P^{\dagger} \otimes I)]=\Tr_A[M]$, the third equality follows from linearity of the trace, and the last equality from $\mathbb{E}_P[P\rho_A^xP^{\dagger}] =I_A/2^{n_A}$.
Hence, $\Chan_B^x(\rho_B^{\mathrm{in}}) $ becomes independent of $x$ and therefore the leakage must be zero as the states can no longer be distinguished.

\section{Equatorial inputs are optimal under $T_1/T_2$ decoherence}
\label{app:optimality_of_plus}
In our experiments, we use a probe prepared in $\ket{+}$. Here, we show that, under the standard $T_1/T_2$ Lindblad model for a superconducting qubit \cite{krantz2019quantum}, this is an optimal choice. Consider the cross-Hamiltonian $H_\times=\zeta Z_a\otimes Z_b$. Since the victim applies no gates, $A$ stays in $\ket{x}$. Here, we neglect relaxation of the victim's qubit during the idle window. This is justified by capping the idle window $\tau$ well below the victim's $T_1$. Any residual decay only reduces the phase gap, so it lowers rather than raises the leakage computed here. This is a modelling idealisation that enables us to calculate the solution explicitly. 
For fixed secret $x$ because $Z_a\ket{x}=(-1)^x\ket{x}$, tracing out the victim's qubit results in the reduced Lindblad master equation:
\begin{equation}
\dot\rho_B^{(x)} = -i(-1)^x\zeta\,[Z_b,\rho_B^{(x)}]
+ \frac{1}{T_1}\Big(\sigma_b^-\rho_B^{(x)}\sigma_b^+
- \tfrac12\Big\{\sigma_b^+\sigma_b^-,\rho_B^{(x)}\Big\}\Big)
+ \frac{\gamma_\phi}{2}\big(Z_b\rho_B^{(x)} Z_b-\rho_B^{(x)}\big),
\qquad \frac1{T_2}=\frac1{2T_1}+\gamma_\phi.
\end{equation}
For a general input state $\ket{\varphi}$, the density operator at time $t$ is
\begin{align*}
\rho_B^{(x)}(t) = {}& \Big(1-\langle 1|\varphi\rangle\langle\varphi|1\rangle\, e^{-t/T_1}\Big)\ket{0}\!\bra{0}
+ \langle 0|\varphi\rangle\langle\varphi|1\rangle\, e^{-t/T_2} e^{-i2(-1)^x\zeta t}\,\ket{0}\!\bra{1} \\
&+ \langle 1|\varphi\rangle\langle\varphi|0\rangle\, e^{-t/T_2} e^{i2(-1)^x\zeta t}\,\ket{1}\!\bra{0}
+ \langle 1|\varphi\rangle\langle\varphi|1\rangle\, e^{-t/T_1}\,\ket{1}\!\bra{1}.
\end{align*}
By the Helstrom bound~\cite{Helstrom1969,Helstrom1976}, the optimal
single-shot success probability for distinguishing $\rho_B^{(0)}(\tau)$ from $\rho_B^{(1)}(\tau)$ is a monotonically increasing function of $\|\rho_B^{(0)}(\tau)-\rho_B^{(1)}(\tau)\|_1$, so maximising the adversary's success means maximising
\begin{equation}
\frac12\big\|\rho_B^{(0)}(\tau)-\rho_B^{(1)}(\tau)\big\|_1
= 2\,\big|\langle0|\varphi\rangle\langle\varphi|1\rangle\big|\, e^{-\tau/T_2}\,\big|\sin(2\zeta\tau)\big|
\end{equation}
over the input state $\ket{\varphi}$. This requires maximising
$|\langle0|\varphi\rangle\langle\varphi|1\rangle|$. Writing
$\ket{\varphi}=\alpha\ket{0}+\beta\ket{1}$ gives
$|\langle0|\varphi\rangle\langle\varphi|1\rangle| = |\alpha||\beta|$,
which is maximised if and only if $|\alpha|=|\beta|=1/\sqrt2$.

The adversary could also entangle its probe with an ancilla. Note conditional spread $\Gamma$ relates to the diamond norm of the difference of the two conditional channels $\mathcal{E}_B^{(0)}$ and $\mathcal{E}_B^{(1)}$ \cite{watrous2018theory}. We have
\[
\mathcal{E}_B^{(x)}=\mathcal N_\tau\circ\mathcal R_x,
\]
where
$\mathcal R_x(\rho)=e^{-i(-1)^x\zeta\tau Z_b}\rho\,e^{i(-1)^x\zeta\tau Z_b}$
is the coherent rotation and $\mathcal N_\tau$ is the $T_1/T_2$ noise
channel over time $\tau$. Writing
$\rho=\tfrac12(I+b_xX+b_yY+b_zZ)$, direct calculation gives
\begin{equation}
(\mathcal R_0-\mathcal R_1)(\rho)=\sin(2\zeta\tau)\,(b_xY-b_yX),
\end{equation}
which has no $I$ or $Z$ component. On traceless operators in the span of $X$ and $Y$ the noise channel acts as multiplication by
$e^{-\tau/T_2}$. Therefore,
\[
\mathcal{E}_B^{(0)}-\mathcal{E}_B^{(1)}=\mathcal N_\tau\circ(\mathcal R_0-\mathcal R_1)
=e^{-\tau/T_2}(\mathcal R_0-\mathcal R_1),
\]
and, as a result,
\begin{equation}
\big\|\mathcal{E}_B^{(0)}-\mathcal{E}_B^{(1)}\big\|_\diamond
=e^{-\tau/T_2}\,\|\mathcal R_0-\mathcal R_1\|_\diamond .
\end{equation}
For two unitary channels with unitaries $U$ and $V$, the diamond-norm
distance satisfies
\[
\frac12\|U(\cdot)U^\dagger-V(\cdot)V^\dagger\|_\diamond
=\sqrt{1-\mathrm{dist}\big(0,\mathrm{conv}\{\lambda_i\}\big)^2},
\]
where $\lambda_i$ are the eigenvalues of $U^\dagger V$ and $\mathrm{dist}$
is the Euclidean distance in the complex plane \cite{kliesch2021theory}.
Here $U^\dagger V=e^{2i\zeta\tau Z_b}$ has eigenvalues $e^{\pm2i\zeta\tau}$,
whose convex hull is a chord at distance $|\cos(2\zeta\tau)|$ from the
origin. Hence $\|\mathcal R_0-\mathcal R_1\|_\diamond=2|\sin(2\zeta\tau)|$
and
\begin{equation}
\Gamma=\frac12\big\|\mathcal{E}_B^{(0)}-\mathcal{E}_B^{(1)}\big\|_\diamond
=e^{-\tau/T_2}\,|\sin(2\zeta\tau)|.
\end{equation}
This equals the unassisted optimum above, so entanglement with an ancilla
gives the adversary no advantage. 

\section{Experimental estimation phase gap and contrast} \label{app:experiment}

For a given idle window $\tau$, we run four circuits pertaining to the combination of the victim qubit prepared in $\ket{0}$ or $\ket{1}$ and the probe qubit read out in $X$ or $Y$. Each circuit is executed independently and run several times to estimate the statistics of the measurements. From the raw counts of each circuit, the corresponding Pauli expectation value is estimated as
\begin{equation}
  \langle O\rangle_x = 2\,\hat P_x(0) - 1, \qquad O\in\{X,Y\},\ x\in\{0,1\},
\end{equation}
where $\hat P_x(0)$ is the fraction of shots returning outcome $0$ for victim state $x$. This gives four
numbers, $\langle X\rangle_0$, $\langle Y\rangle_0$, $\langle X\rangle_1$, $\langle Y\rangle_1$. The phase gap and contrast are then estimated as
\begin{align}
  \Delta\theta &= \theta_0-\theta_1 \ (\mathrm{mod}\ 2\pi,\ \mathrm{wrapped\ to}\ (-\pi,\pi]),
  \\
  C(\tau) &= \tfrac12(c_0+c_1),
  \\
  \Gamma &= C(\tau)\,|\sin(\Delta\theta/2)|,
\end{align}
where, for $x\in\{0,1\}$,
\begin{align}
  \theta_x &= \operatorname{atan2}(\langle Y\rangle_x,\langle X\rangle_x),
  \\
  c_x &= \sqrt{\langle X\rangle_x^2+\langle Y\rangle_x^2}
  .
\end{align}

\paragraph*{Why both victim states are measured.}
For $H_\times=\zeta Z_a\otimes Z_b$, the idealised prediction is $\theta_0=-\theta_1=2\zeta\tau$ exactly, so a single measured branch might seem to fix the other by symmetry. Two effects break this symmetry on real hardware.
First, preparing $\ket{1}$ requires an $X$ gate while preparing $\ket{0}$ does not require any gates. This implies state-preparation errors for $\ket{0}$ and $\ket{1}$ are different. Furthermore, relaxation over the idle window (due to decoherence) carries population from $\ket{1}$ toward $\ket{0}$ at rate $1/T_1$, with no comparable process acting on the $\ket{0}$. Therefore, $c_1$ and $c_0$ decay asymmetrically as a function of $\tau$. Therefore, measuring $\theta_0$, $\theta_1$, $c_0$, $c_1$ independently enables us to account for this asymmetry.

\end{document}